\documentclass{emulateapj}
\usepackage{apjfonts}

\newcommand{\beq}{\begin{equation}}
\newcommand{\beqa}{\begin{eqnarray}}
\newcommand{\eeq}{\end{equation}}
\newcommand{\eeqa}{\end{eqnarray}}
\newcommand{\siml}{\la}

\slugcomment{}

\shorttitle{Unstable GRB photospheres and $e^{\pm}$ lines}
\shortauthors{Ioka et al.}

\begin{document}


\title{
Unstable GRB photospheres
and $e^{\pm}$ annihilation lines
}


\author{Kunihito Ioka\altaffilmark{1}, 
Kohta Murase\altaffilmark{2}, 
Kenji Toma\altaffilmark{1},
Shigehiro Nagataki\altaffilmark{2},
and 
Takashi Nakamura\altaffilmark{1}}


\altaffiltext{1}{
Department of Physics, Kyoto University, 
Kyoto 606-8502, Japan
}
\altaffiltext{2}{
YITP, Kyoto University, Oiwake-cho, Kitashirakawa,
Sakyo-ku, Kyoto, 606-8502, Japan
}

\begin{abstract}
We propose an emission mechanism of prompt gamma-ray bursts (GRBs)
that can reproduce the observed non-thermal spectra
with high radiative efficiencies, $>50\%$.
Internal dissipation below a photosphere can
create a radiation-dominated thermal fireball.
If $e^{\pm}$ pairs outnumber protons,
radiative acceleration of $e^{\pm}$ pairs 
drives the two-stream instabilities
between pairs and protons,
leading to the ``proton sedimentation''
in the accelerating pair frame.
Pairs are continuously shock heated by proton clumps,
scattering the thermal photons into the broken power-law shape,
with the non-thermal energy that is comparable to the proton kinetic energy,
consistent with observations.
Pair photospheres become unstable
around the radius of the progenitor star 
where strong thermalization occurs,
if parameters satisfy
the observed spectral (Yonetoku) relation.
Pair annihilation lines 
are predicted above continua, which could be verified by
GLAST.
\end{abstract}


\section{Introduction and summary}
The emission mechanism of the prompt Gamma-ray bursts (GRBs)
(the most luminous objects in the universe)
is still enigmatic
despite the recent progresses in the Swift era
\citep{meszaros06,zhang07}.
Main issues are (A) the efficiency problem
\citep{ioka06,zhang+07,toma06}
and (B) the cooling problem
\citep{ghi00,meszaros00}.
The first problem (A) is that the GRB radiative efficiency,
defined by the GRB energy divided
by the total energy including the afterglow energy, 
is too high ($>50\%$) to be produced by the internal shocks.
Although the high efficiency may be achieved
by a large dispersion in the Lorentz factor of the outflows
\citep{belo00,koba01},
this would yield smaller correlation coefficient 
for the observed spectral correlations,
such as Amati and Yonetoku relations
\citep{amati06,yonetoku04},
as far as the non-thermal (synchrotron or inverse Compton (IC)) 
processes determine the spectral peak energy.
The recent Swift observations 
make the problem even worse
since the early afterglow energy is smaller than
expected, raising some of GRB efficiencies up to $>90\%$.
The second problem (B) is that the cooling 
time is much shorter than the dynamical time,
making the low-energy spectral slope 
steeper than the observations.

Motivated by these problems,
thermal photosphere models are proposed
where the outflow energy is internally dissipated
and thermalized inside the photosphere
\citep{th94,gc99,meszaros00}.
These models have an advantage
to stabilizing the peak energy,
which is identified with the thermal peak
\citep{th07,rees05}.
Thermal peaks may be actually associated
with up to $30\%$ of long GRBs
\citep{ryde05}.

However a simple photosphere model is not
compatible with observations
that most GRBs are highly non-thermal.
This is one reason that excludes
the original fireball model
\citep{pa86,go86}.
Although Comptonization of the thermal photons 
via magnetic reconnection and/or turbulence
is invoked for non-thermal spectra
\citep{th07,gs07},
the heating mechanism of electrons
is largely uncertain.

In this Letter we propose a possible
scenario to make photospheres non-thermal.
We show that radiatively driven instabilities\footnote{
Our instability is different from that in \cite{wp94}.}
occur in a radiation-dominated photosphere
if $e^{\pm}$ pairs outnumber protons.\footnote{
We only consider protons for baryons for simplicity.}
Since radiation selectively pushes pairs
rather than protons,
the relative velocity between pairs and protons
increases,
driving the two-stream instability,
possibly of the Weibel type (\S~\ref{sec:weibel}).
This produces the small-scale inhomogeneity
of the proton-to-pair ratio,
that could grow up to proton clumps causing shocks with pairs.
Shocked pairs scatter thermal photons into
the observed non-thermal spectra with energy
comparable to the proton kinetic energy, consistent with observations.
The unstable pair photosphere may be common
since the observations including the Yonetoku relation \citep{yonetoku04}
suggest that the instabilities occur
around the radius of the progenitor star where
strong thermalization occurs
(\S~\ref{sec:pair}).
Our model employs the radiation-dominated fireball,
solving (A) the efficiency problem, and
naturally achieves the continuous heating of pairs,
solving (B) the cooling problem (\S~\ref{sec:scenario}).
Pair annihilation lines 
are predicted above continua, 
which could enable GLAST to verify the pair 
photosphere model
\citep{murase07}
(\S~\ref{sec:line}).
We use the unit $k_B=h=1$
and $Q_x=Q/10^x$ in cgs units unless otherwise stated.

\section{Pair photosphere and Yonetoku relation}\label{sec:pair}
Most opacity of the fireball photosphere 
can be provided by $e^{\pm}$ pairs
\citep{rees05,meszaros02,peer04}.
The observed GRB spectrum, if extrapolated,
has a significant fraction of energy 
above the pair production threshold \citep{lith01,baring97}.
So more pairs can be produced
than electrons associated with protons.
(Pairs are $m_p/m_e\sim 2000$ times more abundant
if they have the same energy as protons.)
We assume that internal dissipation such as shocks 
produces pairs via non-thermal
processes and create the pair photosphere.

The plasma and photons are subsequently thermalized
by scatterings under the photosphere.
Assuming that we identify the thermal peak $T_{obs}=\Gamma T$
with the peak energy of GRBs,
we obtain the luminosity $L \sim 4\pi r^2 a T^4 c \Gamma^2$
and hence the comoving size of the photosphere as
\beqa
\ell_p \equiv r/\Gamma
\sim 4 \times 10^8
L_{51}^{1/2} T_{obs,2}^{-2}
\ {\rm cm},
\label{eq:ell}
\eeqa
where $T_{obs,2}=T_{obs}/100{\rm keV}$
and $\Gamma$ is the bulk Lorentz factor.
The optical depth of the pair photosphere is
$\tau_{\gamma e} \sim n_{\pm} \sigma_T \ell_p \sim 1$, which yields
the comoving number density of pairs as
\beqa
n_{\pm} \sim 3 \times 10^{15}
L_{51}^{-1/2}
T_{obs,2}^{2}
\ {\rm cm}^{-3}.
\label{eq:npair}
\eeqa

Our model requires that pairs are accelerated
by radiation, so that the radiation dominates
the pair rest energy,
$a T^4 > n_{\pm} m_e c^2$.
We also require that protons are not accelerated
by radiation exerted on the associated electrons
which is $\sim (n_p/n_\pm) aT^4$
since we are considering $n_{\pm}\gg n_p$ 
and $\tau_{\gamma e} \sim n_{\pm} \sigma_T \ell_p \sim 1$,
so that $(n_p/n_{\pm}) a T^4  < n_p m_p c^2$.
With equations (\ref{eq:ell}) and (\ref{eq:npair}),
these limit the photospheric radius as
\beqa
1 \times 10^{11} T_{obs,2}^{-3/2} L_{51}^{5/8}
\ {\rm cm}
< r < 7 \times 10^{11} T_{obs,2}^{-3/2} L_{51}^{5/8}
\ {\rm cm}.
\eeqa
If we insert the observed Yonetoku relation
\citep{yonetoku04},
\beqa
T_{obs} \propto L^{1/2} 
\quad ({\rm Yonetoku\ relation})
\label{eq:yone}
\eeqa
into the above relation, we have very weak parameter dependence
as  $\propto L^{-1/8}$.
Interestingly the radius in Eq.(3) is comparable to 
the Wolf-Rayet stellar radius
$\sim 2$--$20 R_{\odot}$ \citep{cox00}.
Considering that the strong dissipation
occurs within the progenitor star
via interactions at the jet-star boundary
\citep{zwh04},
most GRBs could satisfy our model assumptions.
Pairs expand to the light speed
within the expansion time $\sim \ell_p/c$
with the comoving acceleration 
\beqa
g \sim \frac{c}{\ell_p/c} \sim 2 \times 10^{12}
L_{51}^{-1/2}
T_{obs,2}^{2}
\ {\rm cm/s}^{2}.
\label{eq:gacc}
\eeqa

For later use we define the mean proton density
$\bar n_p$, which satisfies
$\bar n_p \ll  n_{\pm}$ for the pair photosphere.
In this Letter,  we take $n_{\pm}/\bar n_p = 10^3$,
$T_{obs}=100$ keV, $L=10^{51} {\rm erg/s}$ 
and $\Gamma = 10^3$ for fiducial parameters.
The energy density ratio is then
$a T^4 : \bar n_p m_p c^2 : n_{\pm} m_e c^2 \sim 1 : 0.4 : 0.2$.

\section{Weibel instability between $e^{\pm}$ and protons}\label{sec:weibel}
Radiation selectively pushes pairs rather than protons
because the cross section for scattering is proportional to
the inverse square of mass $\sigma_T \propto m^{-2}$.
Then the relative velocity between pairs and protons
must rise in the initial stage.
Here Coulomb collisions between pairs and protons are
usually negligible.
Collisionless interactions are also absent
if the magnetic fields are weak in the initial fireball.

Then the relative velocity between pairs and protons
drives the two-stream instability,
in particular, of the Weibel type \citep{weibel59}.
Let us first consider the linear stage of the Weibel instability.
The initial distribution function of each component is given by
$f_{j}({\bf v})=({m_j}/{2\pi T_{j}})^{3/2}
e^{-m_j [v_x^2+(v_y-V_j)^2+v_z^2]/2 T_j}$,
where $m_j$ is the mass,
$T_j$ is the temperature,
and $V_j$ is the relative velocity for each component
with $j=e^+, e^-, p$.
Initially the system is non-relativistic.
Linearizing the Vlasov-Maxwell equation, we obtain
the dispersion relation of the pair proton plasma
as
\beqa
c^2 k^2 = \omega^2 - \sum_j \omega_{pj}^2
+ \sum_j \omega_{pj}^2
\frac{T_j + m_j V_j^2}{T_j}
\left[1+\xi_j Z(\xi_j)\right],
\label{eq:disp}
\eeqa
where
$\omega_{pj}^2=4\pi n_j q_j^2/m_j$ is
the plasma frequency,
$Z(\xi)=\pi^{-1/2} 
\int_{-\infty}^{\infty} dx (x-\xi)^{-1} e^{-x^2}$
is the plasma dispersion function,
and $\xi_j=\omega/k (2 T_j/m_j)^{1/2}$ \citep{davidson72}.
Note that we are considering the Lorentz frame
in which the linearized Vlasov-Maxwell equation is
block-diagonal ($D_{xy}=0$),
\beqa
\sum_j \frac{\omega_{pj}^2}{\omega^2}
\left[1+\xi_j Z(\xi_j)\right]
\frac{m_j}{T_j}
\frac{\omega}{k}
V_j=0.
\label{eq:diago}
\eeqa

In the initial stage the relative velocity is small, $m_j V_j^2 \ll T_j$,
so that we may simplify equation (\ref{eq:disp}) since $\xi_j \ll 1$.
With $Z(\xi) \simeq i\pi^{1/2}$ for $\xi \ll 1$,
we find to leading order 
\beqa
\omega \simeq i\left(\frac{2}{\pi}\right)^{1/2}
|k|
\frac{\sum_j \omega_{pj}^2 \frac{m_j V_j^2}{T_j} - c^2 k^2}
{\sum_j \omega_{pj}^2 
\left(\frac{m_j}{T_j}\right)^{1/2}}.
\eeqa
Therefore the plasma is unstable ($\Im \omega > 0$) for $0 \le k^2 \le k_0^2$
where $k_0^2=\sum_j \omega_{pj}^2 m_j V_j^2/c^2 T_j$.
The maximum growth rate is
\beqa
(\Im \omega)_{\max}
\simeq \left(\frac{8}{27\pi}\right)^{1/2}
\frac{
\left[\sum_j \omega_{pj}^2
\frac{m_j V_{j}^2}{T_{j}}
\right]^{3/2}}
{\sum_j \omega_{pj}^2 
\left(\frac{m_j c^2}{T_{j}}\right)^{1/2}
\label{eq:maxg}
}
\eeqa
for 
\beqa
k_{\max}^2=\sum_j \frac{\omega_{pj}^2 m_j V_j^2}{3 c^2 T_j}=k_0^2/3.
\label{eq:kmax}
\eeqa
Equation (\ref{eq:diago}) which characterizes our Lorentz frame
is also reduced to
\beqa
\sum_j \omega_{pj}^2 \frac{m_j}{T_j} V_j
\propto \sum_j \frac{n_j}{T_j} V_j = 0.
\label{eq:diago2}
\eeqa

Let us apply equation (\ref{eq:maxg}) 
to the fireball photosphere.
Noting that (a)
the pair density dominates the proton density,
$n_{+} \sim n_{-} \gg n_p$ (\S~\ref{sec:pair}),
(b) the relative velocity is initially smaller than
the thermal one,
$m_p V_p^2 < T_{+} \sim T_{-} \sim T_p$,
and (c) we have
$-n_{+} V_{+} \sim n_{-} V_{-} \sim n_p V_p$
with equation (\ref{eq:diago2}) and the charge neutrality
$\sum_j q_j n_j V_j=0$,
we find 
$\sum_j \omega_{pj}^2 m_j V_j^2/T_j \sim \omega_{pp}^2 m_p V_p^2/T$
and $\sum_j \omega_{pj}^2 (m_j c^2/T_j)^{1/2} 
\sim 2 \omega_{pe}^2 (m_{e} c^2/T)^{1/2}$.
Then the maximum growth rate in equation (\ref{eq:maxg})
is reduced to 
\beqa
(\Im \omega)_{\max}
&\sim&
\left(\frac{2}{27\pi}\right)^{1/2}
\omega_{pp} 
\frac{\omega_{pp}^2}{\omega_{pe}^2}
\left(\frac{m_p V_p^2}{m_e c^2}\right)^{1/2}
\frac{m_p V_p^2}{T}
\nonumber\\
&\sim& 2\ n_{p,12}^{3/2} n_{-,15}^{-1} T_{-1}^{1/2}
\left(\frac{m_p V_p^2}{T}\right)^{3/2}
\ {\rm s}^{-1},
\label{eq:omax}
\eeqa
where $T_{-1}=T/0.1 {\rm keV}$
and we are assuming $V_p < (T/m_p)^{1/2}$.
Note that the proton current direction is different from the pair one,
and $|V_{+}|\sim |V_{-}| \ll |V_p|$
in the frame (\ref{eq:diago2}).

Comparing the growth rate $(\Im \omega)_{\max}$
in equation (\ref{eq:omax})
with the acceleration rate $\sim g/V_p$ in equation (\ref{eq:gacc}),
we find that the latter is larger
$(\Im \omega)_{\max} < g/V_p$
under the condition
$V_p < (T/m_p)^{1/2} \sim 1 \times 10^7 T_{-1}^{1/2} {\rm cm/s}$.
This means that
the relative velocity $V_p$ overtakes
the thermal one $\sim (T/m_p)^{1/2}$ 
before the Weibel instability develops.

Eventually the Weibel instability saturates
in the velocity region $(T/m_p)^{1/2} < V_p < c$.
Although no non-linear simulation exists so far
for the proton streaming in pairs,
the proton and pair flows will split into current filaments
(cylindrical beam) that is bound by the self-generated magnetic fields
\citep{silva03,nishi03,kato05}.
Since currents with the same (different) direction 
attract (expel) each other,
protons effectively separate from pairs\footnote{
A small fraction of electrons
that preserve the charge neutrality exists in the proton filament.}
and temporarily produce the small-scale inhomogeneity 
of the proton-to-pair ratio.

We note that each current can not separate completely 
if the relative velocity $V_p$
is smaller than the thermal one $\sim (T/m_p)^{1/2}$
because the filament radius $\sim k_{\max}^{-1}$
in equation (\ref{eq:kmax}) is larger than the plasma skin depth,
$r \sim k_{\max}^{-1} > c/\omega_{pp}$,
and hence the current $I=\pi r^2 q n_p V_p$ exceeds
the Alfv$\acute{\rm e}$n current $I_A \equiv m_p c^2 V_p/q$
(the maximum current limited by the self-generated magnetic field)
\citep{alfven39},
i.e., filaments must overlap to reduce the net current.
This saturation is so-called the Alfv$\acute{\rm e}$n limit
\citep{kato05}.
This is not our case since $r < c/\omega_{pp}$ for
$V_p>(T/m_p)^{1/2}$, i.e.,
in the particle limit \citep{kato05}.
Note also that the late evolution is mainly driven by protons
since the kinetic energy of protons dominates 
that of pairs.
The generated magnetic fields are weak
$B^2/8\pi \siml n_p m_p V_p^2 \ll a T^4$ at this stage.

\section{Possible scenario for non-thermalization}
\label{sec:scenario}
\begin{figure}
\plotone{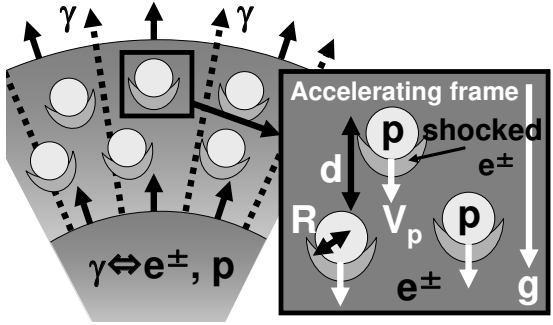}
\caption{
The schematic picture of the unstable GRB photosphere.
Radiation selectively pushes pairs
rather than protons,
driving the two-stream instability between pairs and protons.
This produces the small-scale inhomogeneity
of the proton-to-pair ratio,
that grows up to proton clumps causing shocks with pairs.
In the accelerating pair frame,
particles feel an effective gravity with the acceleration 
$g$ in equation (\ref{eq:gacc}).
Proton clumps fall down and merge, increasing their radii $R$,
infall velocities $V_p$ and vertical separation $d$.
}
\label{fig:sediment}
\end{figure}

We have shown that
protons will temporarily separate from pairs 
on the small-scale possibly
via the Weibel instability.
The subsequent scenario is very speculative
but have attractive features that are worth pursuing.
We consider in the frame comoving with pairs 
(that is almost the same as 
the frame in equation (\ref{eq:diago})).
In this frame, particles feel an effective gravity 
with the acceleration $g$ in equation (\ref{eq:gacc}).
Unlike pairs that interact with radiation, 
protons fall down, which may lead to
the ``proton sedimentation'' in the pair fireball
(see Fig.~\ref{fig:sediment}).

Initially 
protons are trapped by magnetic fields
on the scale of the filament radius.
We model the inhomogeneity by the proton clumps
with the filament size\footnote{
Magnetic clumps are observed in the 
particle simulations,
though not applicable to our pair proton case.
\citep{chang07}.}
($R \sim c/\omega_{pp}$).
Since a clump consists of many particles,
the center-of-mass thermal motion
is much less than the infall velocity.
Also the acceleration time is 
comparable to the growth time of the instability,
so that the clumps continue falling down
and collide with other clumps within
$t_m \sim R/\delta V_p \sim 2 \times 10^{-6} n_{p,12}^{1/2} V_{p,7}^{-1}$ s
where $\delta V_p \sim V_p$ is
the relative velocity of clumps.
Since the Larmor radius is comparable to the clump size,
the fluid approximation begins to be valid.
Then, we may apply the Stokes' law to estimate 
the terminal infall velocity as
$V_p \propto n_p R^2$.
Since the number density of clumps
is $N_c \propto n_p^{-1} R^{-3}$ (mass conservation)
and the cross section for the collision is
$\sigma_c \propto R^2$,
the merger timescale is
\beqa
t_m \sim (N_c \sigma_c \delta V_p)^{-1}
\propto R^{-1}.
\eeqa
This leads to runaway growth of clumps
because the merger timescale is shorter for larger clumps.

After the terminal velocity exceeds the sonic speed
of pairs $V_p > c_{s,\pm}$,
we can not use the Stokes' law but instead the momentum balance.
Since the pair mass colliding with a proton clump during $\Delta t$
is $\Delta M \sim m_e n_{\pm} R^2 V_p \Delta t$,
the momentum conservation gives
the velocity change $\Delta V_p \sim V_p \Delta M/m_p n_p R^3$.
Equating $\Delta V_p/\Delta t$ with the acceleration $g$,
we find the terminal velocity as
\beqa
V_p \sim \left(\frac{m_p n_p g R}{m_e n_{\pm}}\right)^{1/2}.
\eeqa
With $g \sim c^2/\ell_p$ in equation (\ref{eq:gacc}),
the terminal velocity reaches the light speed $V_p \sim c$
when the ratio of clumps to the system size is
$R/\ell_p \sim m_e n_{\pm}/m_p n_p < 1$.
So the final separation between clumps 
in the vertical direction is about
\beqa
d \sim \frac{n_p}{\bar n_p} R \sim \ell_p \frac{n_{\pm} m_e}{\bar n_p m_p},
\label{eq:d}
\eeqa
since the swept mass is $R^2 d \bar n_p \sim R^3 n_p$.

Collisionless shocks arise
as the relative velocity between proton clumps and pairs approaches
the light speed.
Pairs are shock heated and accelerated
to a power-law distribution
$N(\gamma_{\pm}) d\gamma_{\pm}
\propto \gamma_{\pm}^{-p} d\gamma_{\pm}$
for the pair Lorentz factor $\gamma_{\pm} \ge \gamma_m \sim 1$,
where  $\gamma_m$ is the minimum Lorentz factor.
Shocked pairs are pushed sideways and then again shocked by other clumps
(see Fig.~\ref{fig:sediment}).
The timescale between shocks is about 
$\sim d/c$ in equation (\ref{eq:d}),
and interestingly
this is comparable to
the IC cooling timescale,
\beqa
{\cal R} \equiv \frac{t_{c}}{d/c} 
\sim \frac{\gamma_{\pm} m_e c^2}
{c \sigma_T U_\gamma \gamma_{\pm}^2}
\frac{c}{d}
\sim 
\frac{\bar n_p m_p c^2}{\gamma_{\pm} U_{\gamma}},
\eeqa
for $\gamma_{\pm} \sim 1$
if the proton energy is comparable to
the radiation energy $\bar n_p m_p c^2 \sim U_\gamma \equiv aT^4$
(as our parameters),
where we use $\tau_{\gamma e} \sim n_{\pm} \sigma_T \ell_p \sim 1$
in the second equality.
{\it Therefore the clumps automatically have the right separation
for the continuous heating of pairs.}
Since a photon is scattered by $\gamma_{\pm} \sim 1$ pairs
once on average for $\tau_{\gamma e} \sim 1$,
scattered photons have the energy density 
\beqa
{\cal R} U_\gamma \sim 
\bar n_p m_p c^2.
\eeqa
{\it Thus, the non-thermal component has comparable energy to the proton
kinetic energy, consistent with observations.}

\begin{figure}
\plotone{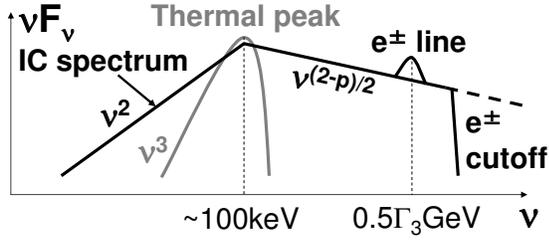}
\caption{
The schematic picture of the GRB spectrum 
in our unstable photosphere model.
Shocked pairs scatter thermal photons into
the observed broken power-law spectra with energy
comparable to the proton kinetic energy, consistent with observations.
The pair annihilation line is predicted above continua
in the pair photosphere model,
which is blueshifted to $\nu
\sim 0.5 \Gamma_{3}$ GeV 
and broadened by the order-of-unity distribution of 
the Lorentz factor on the photosphere.
At higher energy, a spectral cutoff arises due to
the pair creation.
A closure relation exists  
between observable quantities of
lines and cutoffs \citep{murase07},
which may be verified by GLAST.
}
\label{fig:spec}
\end{figure}

Neglecting multiple scatterings since $\tau_{\gamma e} \sim 1$,
we may calculate the observed IC spectrum as \citep{se01}
\beqa
F_{\nu}=\left\{
\begin{array}{ll}
(\nu/T_{obs}) F_{\nu,\max},& \nu < T_{obs},\\
(\nu/T_{obs})^{-p/2} F_{\nu,\max},& T_{obs} < \nu,
\end{array}\right.
\eeqa
that also peaks at the thermal peak $\nu \sim T_{obs}$ for $p>2$
because $\gamma_m \sim 1$
(see Figure~\ref{fig:spec}).
The ratio of the IC peak flux to the thermal one
is $F_{\nu,\max}/F_{\nu,\max}^{BB} \sim 
\bar n_p m_p c^2/ U_\gamma$
($\sim 1$ for our parameters).
At high frequencies $\nu > T_{obs}$,
the IC spectrum has a cooling spectrum $\propto \nu^{-p/2}$
since the cooling is faster than the heating,
$t_c < d/c$, for $\gamma_{\pm}>1$.
At low frequencies $\nu < T_{obs}$,
the Rayleigh-Jeans spectrum $\propto \nu^2$
is modified to $\propto \nu$ by down-scatterings.
Therefore the total (IC plus thermal) spectrum
is roughly a broken power-law
peaking at $\nu \sim T_{obs}$
and resembles the observed GRB spectrum although the
low energy photon index is slightly harder.

\section{Prediction of $e^{\pm}$ annihilation lines}\label{sec:line}
We can prove that the pair photosphere
accompanies the pair annihilation line
that is detectable above the power-law continuum if $\gamma_{\pm} \sim 1$.
Let us assume that the continuum exceeds the annihilation line,
$n_{\gamma,c}>n_{\gamma,l}$, where
$n_{\gamma,c}$ ($n_{\gamma,l}$) is 
the number density of continuum (line) photons at the line energy.
On the pair photosphere, $n_{\pm} \sigma_T l \sim 1$,
we have $n_{\gamma,l} \sim n_{\pm}$
because the pairs annihilate once on average
in the dynamical time,
$\dot n_{\pm} \ell_p/c
\sim -n_{\pm}^2 \sigma_T \ell_p 
\sim -n_{\pm}$ \citep{sven87}.
Then the optical depth to the pair production is larger than unity,
\beqa
\tau_{\gamma \gamma}
\sim n_{\gamma,c} \sigma_T \ell_p > n_{\gamma,l} \sigma_T \ell_p
\sim n_{\pm} \sigma_T \ell_p \sim 1,
\eeqa
and hence the continuum photons 
create more pairs than the existing pairs.
However this contradicts with the definition of the pair photosphere.
Therefore the assumption $n_{\gamma,c}>n_{\gamma,l}$
is wrong and the annihilation line should stand out of
the continuum.
As is clear from the above proof,
the line flux is comparable to the continuum one
when pairs are created from the continuum photons.

The lines are blueshifted to $\nu
\sim 0.5 \Gamma_{3}$ GeV \citep{peer06}
and broadened by the order-of-unity distribution of 
the Lorentz factor on the photosphere.
At higher energy, a spectral cutoff arises due to
the pair creation \citep{asano07,baring97,lith01,soeb04}.
A closure relation exists  
between observable quantities of
lines and cutoffs for the pair photosphere
\citep{murase07},
which may be verified by GLAST.

\section{Discussions}
The Yonetoku relation (\ref{eq:yone}) can be derived
from equation (\ref{eq:ell})
if the photosphere appears at the stellar radius
$r \sim$ const., as suggested in \S~\ref{sec:pair},
and the outflow rate of slow mass satisfies
$\dot M_s \propto \Gamma_s$
because the Lorentz factor after two mass collision
is about
$\Gamma \sim (L \Gamma_s/\dot M_s c^2)^{1/2} \propto L^{1/2}$.
We can test $\Gamma \propto L^{1/2}$ 
since $\Gamma$ is determined by the annihilation lines.

Our model predicts
(a) (non-thermal energy) $\sim$ (proton energy) $\sim$ (afterglow energy)
and hence thermal peaks may outstand for
GRBs with $>90\%$ efficiencies.
(b) Low energy index that is comparable to the observed
hardest one.
Superpositions of pulses
or incomplete thermalization may produce softer index.
(c) Dim optical flashes from reverse shocks \citep{li03},
consistent with recent observations.
(d) Polarization in the non-thermal component
(except for the thermal peak) produced by IC scatterings.
(e) High energy cosmic rays and neutrinos
since protons might be also accelerated at shocks.
Particle simulations with pairs and protons 
are needed to verify our model.
Other plasma instabilities,
such as the electrostatic two-stream instability
or the drift-wave instability with the initial
magnetic fields,
may also generate the initial proton inhomogeneity.

\acknowledgments
We thank K.~Asano and T.~Kato for useful comments. 
This work is supported in part 
by Grant-in-Aid for the 21st Century COE
``Center for Diversity and Universality in Physics''
from the Ministry of Education, Culture, Sports, Science and Technology
(MEXT) of Japan and by
the Grant-in-Aid from the 
Ministry of Education, Culture, Sports, Science and Technology
(MEXT) of Japan, No.18740147 (K.I.),
No.19104006, No.19740139 (S.N.)
and No.19540283, No.19047004 (T.N.).

\end{document}